\documentclass{mem}
\usepackage{natbib}\usepackage{txfonts}\usepackage{balance}
\usepackage{graphicx}
\usepackage[a4paper,breaklinks,dvipdfm]{hyperref}
\usepackage{color}
\idline{}{1}
\begin{document}
\def\teff{$T\rm_{eff }$}
\def\kms{$\mathrm {km s}^{-1}$}
\def\lsim{\;\raise0.3ex\hbox{$<$\kern-0.75em\raise-1.1ex\hbox{$\sim$}}\;}
\def\gsim{\;\raise0.3ex\hbox{$>$\kern-0.75em\raise-1.1ex\hbox{$\sim$}}\;}
\newcommand{\red}{\textcolor{red}}
\newcommand{\blue}{\textcolor{blue}}
\newcommand{\green}{\textcolor{green}}
\newcommand{\DSA}{diffusive shock acceleration}
\newcommand{\MFA}{magnetic field amplification}
\newcommand{\Msun}{\mbox{$M_{\odot}\;$}}
\newcommand{\heatpar}{\alpha_H}
\def\lsim{\;\raise0.3ex\hbox{$<$\kern-0.75em\raise-1.1ex\hbox{$\sim$}}\;}
\def\gsim{\;\raise0.3ex\hbox{$>$\kern-0.75em\raise-1.1ex\hbox{$\sim$}}\;}
\def\alf{Alfv\'en }
\def \cms {\rm ~cm~s^{-1}}
\def\cmc{\rm ~cm^{-3}}
\def\kms{\rm ~km~s^{-1}}
\def\ml{~\Msun ~\rm yr^{-1}}
\def\mll{\Msun ~\rm yr^{-1}}
\def\etal{{ et al. }}
\def\cmc{\rm ~cm^{-3}}
\def\diff{\rm ~cm^2~s^{-1}}
\def \kms {\rm ~km~s^{-1}}
\def \cmsec {\rm ~cm~s^{-1}}
\def\ergs{\rm ~erg~s^{-1}}
\def\phfl{\rm ~ph~cm^{-2}~s^{-1}~keV^{-1}}
\def\lfl{\rm ~ph~cm^{-2}~s^{-1}}
\def\enf{\rm ~erg~cm^{-2}~s^{-1}}
\def\arcmin{\hbox{$^\prime$}}
\def\arcsec{\hbox{$^{\prime\prime}$}}
\def\efl{\hbox{erg cm$^{-2}$ s$^{-1}$}}
\def\kpc{{\rm\thinspace kpc}}
\def\mpc{{\rm\thinspace Mpc}}
\def\pasj{PASJ}
\title{
    Particle acceleration at supernova shocks in young stellar clusters}

   \subtitle{}

\author{
A.M. \,Bykov\inst{1} \and P.E.\,Gladilin\inst{1} \and
S.M.\,Osipov\inst{1,2}
          }

  \offprints{A.M. Bykov}

\institute{A.F.Ioffe Physical-Technical Institute, St. Petersburg,
Russia
  \email{byk@astro.ioffe.ru} \and State Politechnical University, St.\
Petersburg, Russia}

\authorrunning{A.M.Bykov et al}

\titlerunning{The sources of galactic cosmic rays}

\abstract{ We briefly discuss models of energetic particle
acceleration by supernova shock in active starforming regions at
different stages of their evolution. Strong shocks may strongly
amplify magnetic fields due to cosmic ray driven instabilities. We
discuss the magnetic field amplification emphasizing the role of the
long-wavelength instabilities. Supernova shock propagating in the
vicinity of a powerful stellar wind in a young stellar cluster is
argued to increase the maximal CR energies at a given evolution
stage of supernova remnant (SNR) and can convert a sizeable fraction
of the kinetic energy release into energetic particles.}
\maketitle{}

\section{Introduction}
The observed spectra of Galactic cosmic rays (CRs) are shaped by two
basic processes - the acceleration in the sources and the subsequent
propagation in cosmic magnetic fields and radiation fields. A
transition from galactic to extragalactic cosmic rays is expected to
occur somewhere between $10^{17}$~eV and $10^{19}$~eV
\citep[e.g.][]{hillas05,aharonianea11}. A preponderance of evidence
suggests that the particle acceleration mechanism most likely
responsible is \DSA\ (DSA) \citep[e.g.,][]{BE87,JE91,md01}.

An important question for DSA and CR origin has always centered
around the maximum particle energy a given shock can produce. For a
shock of a given size, age, and magnetic field geometry, the maximum
CR energy depends mainly on the power in the longest wavelength
turbulence. The weakly anisotropic distribution of accelerated
particles, i.e., CRs is considered in \S~\ref{mf1} as an agent
producing this turbulence in a symbiotic relationship where the
magnetic turbulence required to accelerate the CRs is created by the
accelerated CRs themselves.

Other important issue of the high energy CR acceleration models is
the environment where the supernova exploded.  Some models of CRs
acceleration by SNRs in active star forming regions are briefly
discussed in \S~\ref{SB}. OB-associations and young globular
clusters are observed both in the Milky Way and LMC. The study of
the stellar content of the galactic object  Cygnus OB2  by
\citet{Knoedlseder00} have revealed that the number of OB member
stars can be estimated as large as 2600 $\pm$ 400, while the number
of O stars amounts to 120 $\pm$ 20. The high number of stellar X-ray
sources detected with {\sl Chandra} by \citet{wd09} confirmed the
status of Cygnus OB2 as one of the most massive SFRs in the Milky
Way. Given the apparently compact size of Cygnus OB2 one may expect
a number of massive stars with strong winds to be in a close
proximity with less than 10 pc separation in addition to the well
known colliding-wind binaries that are expected to be particle
accelerators \citep[e.g.][]{eu93}. Another very compact young
stellar cluster Westerlund 2 containing more than dozen O stars was
found in the Carina region with estimated age to be younger than 4
Myrs, so the most massive stars are expected to explode there within
the next few Myrs. OB associations with supernova explosions are
creating superbubbles (SBs). \citet{bambaea04} discovered both
thermal and nonthermal X-rays from the shells of the SB 30 Dor C in
the LMC. The X-ray morphology was reported as a nearly circular
shell with a radius of ~40 pc, which is bright on the northern and
western sides. \citet{maddoxea09} analyzed {\sl Suzaku} observations
of the SB around the OB association LH9 in the HII complex N11 in
the Large Magellanic Cloud. Their X-ray spectral analysis revealed
that the hard X-ray emission ($>2$ keV) requires a hard nonthermal
power-law component. The energy budget analysis for N11 using the
known stellar content of LH9 indicated that the observed thermal and
kinetic energy in the SB is only half of the expected mechanical
energy injected by stars, consistent with the expectation of SB
models with efficient CR acceleration \citep[e.g.][]{bykov01,bb08}.
Diffuse X-ray emission was detected from many sites of massive star
formation: the Carina Nebula, M17, 30 Doradus, NGC 3576, NGC 3603,
and others \citep[e.g.][]{Townsley11}. The H.E.S.S. telescope
detected high-energy gamma rays from starburst galaxy NGC 253
supporting the ideas of efficient CR acceleration in active
starforming regions \citep{ngc253}.

\begin{figure*}[t!]
\resizebox{\hsize}{!}{\includegraphics[clip=true]{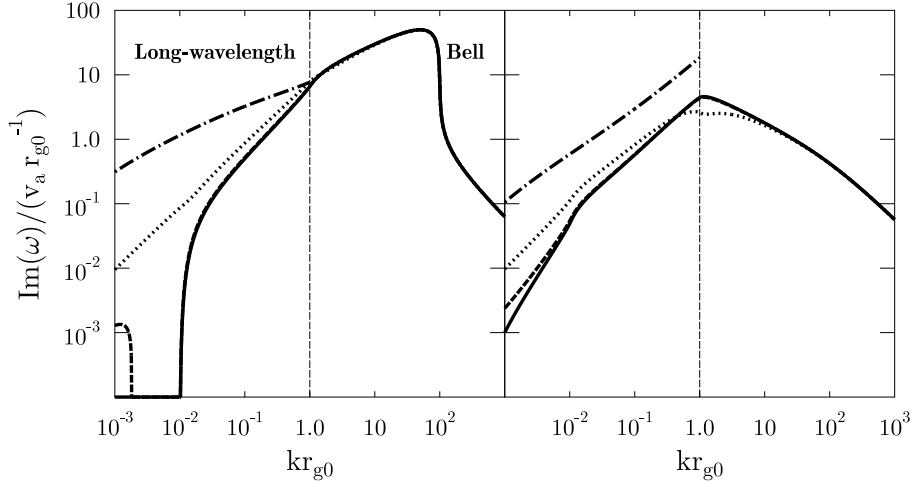}}
\caption{\footnotesize The two panels show the growth rates for two
opposite circular polarization modes (shown in left and right
panels) of the parallel propagating modes as a function of the
wavenumber. The CR resonant streaming instability and the Bell
instability The dot-dashed curves effect of short-scale turbulence
on the long-wavelength instability. The model parameters are $\Delta
= 100$, $\beta_s = 0.01$ and $\alpha = 4.0$. For comparison, the
dot-dashed curves are calculated for the model of the ponderomotive
CR current instability in the presence of the short-scale turbulence
\citep[][]{boe11}. The dotted curves illustrate the maximal case of
the CR-firehose instability (i.e., $\delta \propto \beta_s$). The
dashed curve corresponds to  $\delta = 5\,\beta_s^2$.} \label{mf}
\end{figure*}

\section{Magnetic field amplification in CR acceleration
sources}\label{mf1} Fast and efficient CR acceleration by Fermi
mechanism requires that particles are multiply scattered by magnetic
fluctuations in the acceleration source (e.g. shock). The magnitude
of the required magnetic fluctuations is substantially higher than
the ambient magnetic turbulence forcing a bootstrap scenario where
the accelerated particles amplify the turbulence required for their
acceleration. The study of turbulence generation associated with CRs
and DSA has a long history. Magnetic field amplification due to the
resonant cosmic-ray streaming instability was studied in the context
of galactic cosmic-ray origin and propagation since the 1960s
\citep[see e.g.][]{kc71,wentzel74,achterberg81,berea90, zweibel03}.
It was proposed by \citet{bell78} as a source of magnetic turbulence
in the test particle DSA scenario, and nonlinear models of DSA
including CR-driven instabilities and magnetic field amplification
were investigated by \citet{bell04,ab06,zp08},\citet{vbe08,vbe09}
and \citet{rkd09}.

It is instructive to summarize the  growth rates for magnetic
instabilities that the quasi-linear theory predicts for weakly
anisotropic CR distributions of the form
\begin{equation}\label{distrF0}
f(\bf{p})=\frac{N}{4\pi}\left[1+3\beta_{s}\mu+\frac{\delta}{2}\left(3\mu^{2}-1\right)\right],
\end{equation}
where  $\theta$ - particle pitch-angle, $\mu=\cos\theta$ ,
$\beta_s=u_{s}/c$, with the flow (shock) velocity $u_{s}$,
$\delta(p)$ - is the magnitude of the second harmonic anisotropy. We
included here the second harmonic anisotropy to illustrate the
effect of the CR-firehose instability on the magnetic field
amplification. The substantial anisotropy of the second harmonics
can be expected in more complex plasma flows than just an isolated
shock. The isotropic part of the distribution function is a
power-law of the form
\begin{equation}\label{spektrNp}
N\left(p\right)=\frac{n_{cr}\left(\alpha-3\right)p_{0}^{\left(\alpha-3\right)}}{\left[1-\Delta^{\alpha-3}\right]p^{\alpha}},
\end{equation}
where the CR momenta $p_{0}\leq p\leq p_{m}$, $\Delta =
\frac{p_{m}}{p_{0}}$,  $\alpha$ - is the CR power law index, and
$n_{cr}$- is the CR number density. The CR mean free path $\Lambda =
\eta r_{g}(p)$, where the particle gyroradius is $r_{g}(p)
=\frac{cp}{eB_{0}}$ and $\eta \geq 1$.  Then, following the standard
linear analysis of the kinetic equation {\sl in the intermediate
regime} $\eta^{-1} < x_0< 1$ \citep[see e.g.][]{boe11} one may get
the following dispersion relation:
\begin{equation}\label{dispers}
\frac{\omega^{2}}{v_{a}^{2}k^{2}}=\left[1\mp
\frac{k_{0}}{k}\left\{A_{0}-1+\frac{\delta}{\beta_{s}}A_{1}\right\}\right].
\end{equation}
\begin{equation}\label{koeffA}
A_{0,1}\left(x_{0},x_{m}\right)=\int_{p_{0}}^{p_{m}}\sigma_{0,1}\left(p\right)N\left(p\right)p^{2}dp
\end{equation}
\begin{equation}\label{sigma0Int}
\sigma_{0}\left(p\right)=\frac{3}{4}\int_{-1}^{1}\frac{\left(1-\mu^{2}\right)}{1\mp
x\mu}d\mu,
\end{equation}
\begin{equation}\label{sigma1Int}
\sigma_{1}\left(p\right)=\frac{3}{4}\int_{-1}^{1}\frac{\left(1-\mu^{2}\right)\mu}{1\mp
x\mu}d\mu
\end{equation}
where $\displaystyle
k_{0}=\frac{4\pi}{c}\frac{en_{cr}u_{s}}{B_{0}}$, $x=k r_{g}(p)$ ,
$x_{0}=k r_{g}(p_0)$ , $x_{m}= k r_{g}(p_m)$. The signs $\pm$
correspond to the two opposite circularly polarized modes under
investigation. The second term in the right hand site in corresponds
to the fast growing mode discovered by \citet[][]{bell04} for
$x_{0}>1$. The instability results in amplification of a mode with
wavenumbers $\displaystyle k_{0}>k>\frac{1}{r_{g0}}$. For $x_{0}<1$
the term in Eq.~\ref{dispers} is responsible for the well known
resonant streaming instability \citep[e.g.][and the references
therein]{achterberg81,zweibel03,plm06,mlp06,ab09}. The last term in
the r.h.s. of Eq.~\ref{dispers} described the CR firehose
instability. In the long-wavelength regime  $x_{m} \ll 1$  a
simplified form of Eq.~\ref{dispers} can be derived
\begin{equation}\label{dispers2asimpX0}
\frac{\omega^{2}}{v_{a}^{2}k^{2}}=1\mp
\frac{k_{0}r_{g0}}{5}\left[x_{m}\pm\frac{\delta\beta_{s}^{-1}\ln\Delta}{\left(1-\Delta^{-1}\right)}\right].
\end{equation}
The growth rates of the unstable modes from the dispersion relation
Eq.~\ref{dispers} are shown in Figure~\ref{mf}. The growth rates of
the resonant and the Bell instabilities are shown by solid lines in
Figure~\ref{mf}. Since $x_{m} \propto k$ at $x_{m} \sim 1$, the
growth rate $Im(\omega) \propto k^{3/2}$ -- the long-wavelength
regime pointed out by \citet{sb11} (for $\Delta =100$). At longer
wavelengths however the second term in the square brackets in
Eq.~(\ref{dispers2asimpX0}) dominate and the CR-firehose type
instability appear at $x_{m} << 1$. The CR-firehose instability rate
is indicated by the dashed line at the left panel in
Figure~\ref{mf}. The CR-firehose instability may even dominate the
regime $Im(\omega) \propto k^{3/2}$ in the case of the second
harmonic comparable with the first harmonic (CR current) of the
anisotropy, i.e. $\delta \propto \beta_s$.  The CR-firehose growth
rate in that regime is shown by the dotted curve in Figure~\ref{mf}.
That likely requires a more complicated structure of the magnetic
field in the flow. The flow in the vicinity of the SNR colliding
with a powerful stellar wind \citep[see e.g.][]{velazquesea03}
discussed below as the PeV CR acceleration site may be relevant. To
derive the growth rates of the modes in the long-wavelength regime
$k \Lambda <1$ (i.e. $x_0 <  \eta^{-1}$) the collisionless kinetic
equation approach discussed above is not appropriate. One should
average the kinetic equation for the relativistic particles, the
equations of the bulk plasma motions and the induction equation over
the ensemble of the short scale fluctuations produced by CR
instabilities in the collisionless regime e.g. by the fast Bell
instability. In the presence of the short-scale fluctuations, the
momentum exchange between the CRs and the flow in the hydrodynamic
regime, results in a ponderomotive force that depends on the CR
current in the mean-field momentum equation of bulk plasma
\citep[][]{boe11}. As a result, there exist transverse growing modes
with wavevectors along the initial magnetic field with growth rates
that are proportional to the turbulent coefficients determined by
the short scale fluctuations. The magnetic field amplification in
that regime only weakly depends on the shock velocity [as it follows
from Eq.~(48) in \citet{boe11}], that is important for the evolution
of the maximal energy of CRs accelerated by DSA.

\begin{figure}[t!]
\resizebox{\hsize}{!}{\includegraphics[scale=0.6]{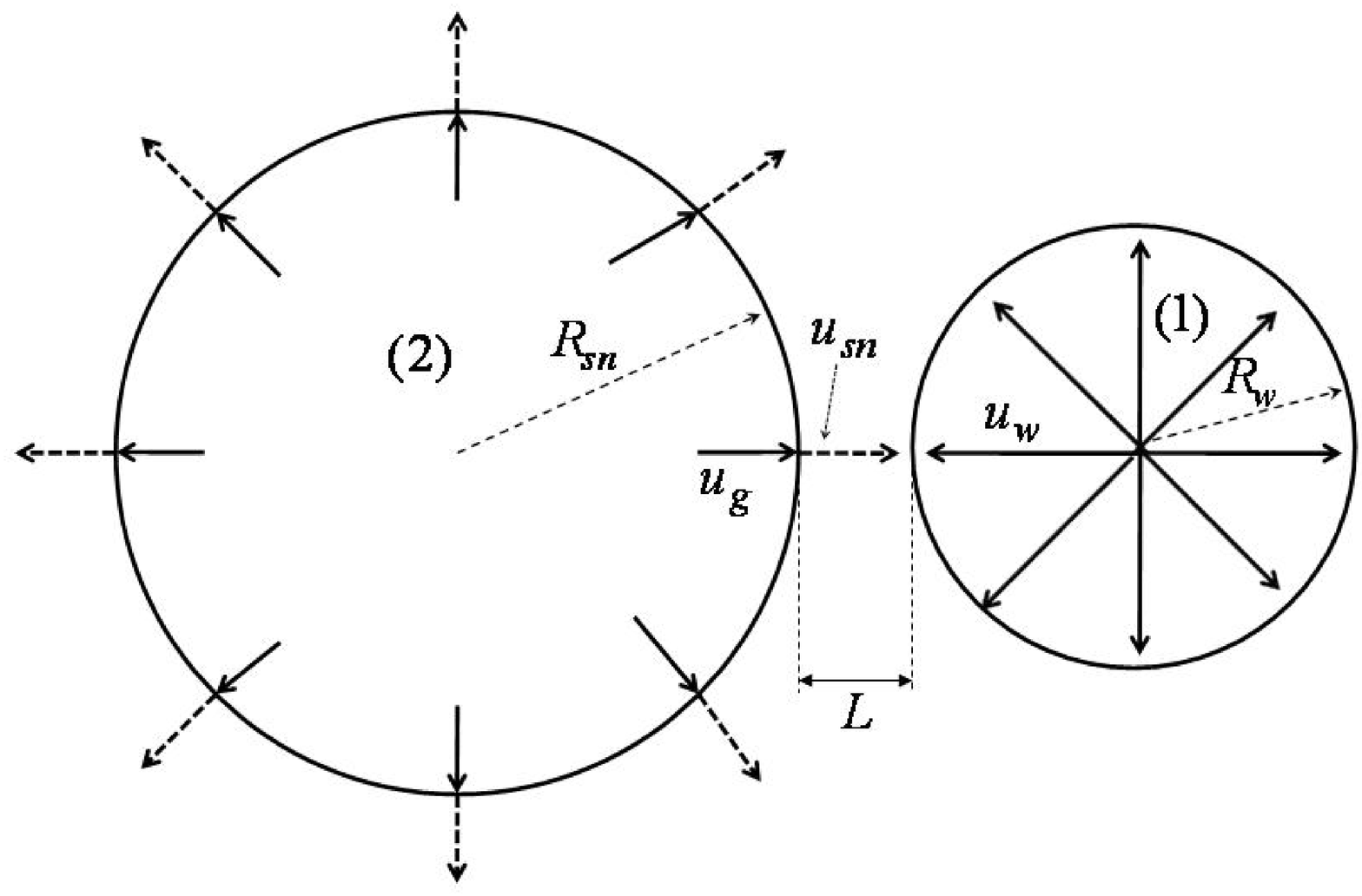}}
\caption{\footnotesize Simplified view of the expanding supernova
remnant approaching a powerful stellar wind. The configuration is
shown to be favorable for 10$^{15}$ eV regime CR proton acceleration
at a few thousands years old SNR in SFRs. $R_w$ and $R_{sn}$ are the
radii of the stellar wind termination shock and SNR forward shock,
respectively. $u_w$ is velocity of the stellar wind, $u_{sn}$ is the
SNR forward shock velocity, $L$ is the distance between the two
flows.} \label{snr_wind}
\end{figure}

\begin{figure}[t!]
\resizebox{\hsize}{!}{\includegraphics[bb= 0 30 450 300]{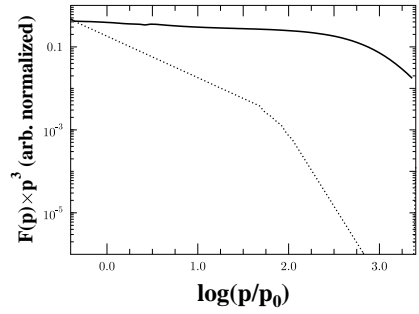}}
\caption{\footnotesize CR particle spectrum (solid line) from the
time-dependent DSA simulation of the SNR forward shock approaching
the O-star wind termination shock [see Figure \ref{snr_wind}]. For
the comparison the CR spectrum at solitary shock is shown as the
dotted line curve.} \label{spectra}
\end{figure}

\section{Particle acceleration by a supernova remnant in OB
association}\label{SB} The maximal energies of CRs accelerated by
supernova shocks with magnetic field amplification can reach the
knee region of the GCR spectrum
\citep[e.g.][]{BE87,bell04,bv07,cba11}. Moreover, it was recently
argued by \citet{pzs10} that the maximum energy of accelerated
particles differs strongly for different supernova types. The energy
may exceed $10^{17}$ eV for protons accelerated in an isolated Type
IIb SNR. The highest energies CRs are thought to be accelerated
before the Sedov stage in isolated SNRs.

For supernova exploded in a young stellar cluster the presence of a
nearby strong stellar wind allow to increase the maximum energy of
accelerated particles comparing to that in an isolated SNR. The SNRs
can produce the high energy CRs at the Sedov evolution stage.

The most massive O stars begin to explode as core collapsed
supernovae about a few million years after the formation of a young
stellar cluster. The young stars with masses above $\sim 16\,\Msun$
(of B0~V type and earlier) are thought to create hot low-density
bubbles and HII regions with radii $\sim 10$\,pc surrounded by a
massive shell of matter swept up from the parent cloud by the
stellar wind.
In this case, a strong supernova shock propagates for a few thousand
years in tenuous circumstellar matter with a velocity well above
$10^3$\,km/s before reaching the dense massive shell. The blast wave
of the SNR is expected to accelerate the wind material producing
ultra-relativistic ions and electrons. The accelerated particles
escape from the SNR forward shock presumably at the highest energy
regime \citep[e.g.][]{pzs10,gabici11,eb11} and can reach the
termination shock of the stellar wind of a nearby massive star. We
modeled the energetic particle acceleration in the region where the
expanding supernova shell is approaching a powerful stellar wind of
a young massive star as it is illustrated in Figure~\ref{snr_wind}.
At the evolution stage where the mean free path of the highest
energy CR is comparable to the distance between the two shocks $L$
the system is characterized by an unusually hard spectral energy
distribution illustrated by the solid line in Figure~\ref{spectra}.
The stage can last more than 1,000 yrs. For comparison we showed
with the dotted line the CR spectrum accelerated by an isolated SNR
shock of the same age. The case of Bohm diffusion was simulated with
$D(p) \propto p$.

A simple analytical kinetic model of particle acceleration by the
Fermi mechanism in the converging flows carrying fluctuating
magnetic fields amplified by anisotropic CR distributions can be
considered. It is a model of high energy CR acceleration by a SNR
expanding in a compact OB-association. At some expansion phase the
distance $L$ between the SNR approaching the stellar wind shock is
less than the mean free path of the highest energy CR particle in
the SNR shock precursor. Then CR distribution function around the
shocks $(i = 1,2)$ can be approximated as
\begin{equation}\label{SolveDC}
\begin{array}{l}
 f_i \left( {x,p,t} \right)= A p^{-3}\exp \left( {-\frac{u_i }{D_i }\left| x \right|}
\right)\times \\
 \times H\left( {p-p_0 } \right)H\left( {t-\tau _a } \right)
 \end{array}
\end{equation}
where the highest energy CR acceleration time
\begin{equation}\label{TauAc}
\tau _a =\int\limits_{p_0 }^p {\frac{3}{\left( {u_1 +u_2 }
\right)}\left( {\frac{D_1 }{u_1 }+\frac{D_2 }{u_2 }} \right)}
\frac{dp}{p}.
\end{equation}

The acceleration mechanism can provide efficient creation of a
nonthermal particle population with a very hard energy spectrum,
containing a substantial part of the kinetic energy released by the
supernova. The high energy nonthermal emission of the sources is
characterized by a very hard spectral energy distribution peaked at
the maximal photon energies and have the apparent properties of the
"dark accelerators". The TeV source in the vicinity of the
$\gamma$-Cygni SNR can be belong to the class.


At the later stage of the young stellar cluster evolution multiple
supernova explosions with great energy release in the form of shock
waves inside the superbubbles are argued as a favorable site of
nonthermal particle acceleration. The collective effect of both
stellar winds of massive stars and core collapsed supernovae as
particle acceleration agents were discussed by
\citet[][]{bt01,parizotea04,td07,fm10}.

\begin{acknowledgements}
A.M.B. thanks A.Marcowith for the very interesting CRISM meeting and
ISSI (Bern) team. The authors were supported in part by
 the Russian government grant
11.G34.31.0001 to Sankt-Petersburg State Politechnical University,
by the RAS Presidium and OFN Programs, by the RFBR grant
11-02-12082. They performed the simulations at the Joint
Supercomputing Centre (JSCC RAS) and the Supercomputing Centre at
Ioffe Institute, St. Petersburg.
\end{acknowledgements}




\end{document}